\documentclass{aa}
\usepackage{natbib}
\usepackage{graphicx}
\bibpunct{(}{)}{;}{a}{}{,} 

\begin{document}

\title{Revealing the fastest component of the DG Tau outflow through X-rays}

\author{H.M. G\"unther\inst{1, 2} \and S. P. Matt\inst{2} \and Z.-Y. Li\inst{2}}

\institute{Hamburger Sternwarte, Universit\"at Hamburg, Gojenbergsweg
  112, 21029 Hamburg, Germany\\ \email{moritz.guenther@hs.uni-hamburg.de} \and Department of Astronomy, University
  of Virginia, P.O. Box 400325, Charlottesville, VA 22904, USA}

\date{Received 2 September 2008 / Accepted 4 November 2008}

\abstract{Some T Tauri stars show a peculiar X-ray
  spectrum that can be modelled by two components with different
  absorbing column densities.}  {We seek to explain the soft X-ray
  component in DG~Tau, the best studied of these sources, with an
  outflow model, taking observations at other
  wavelengths into consideration.}  {We constrain the outflow properties through spectral
  fitting and employ simple semi-analytical formulae to describe
  properties of a shock wave that heats up the X-ray emitting region.}
  {The X-ray emission is consistent with its arising
  from the fastest and innermost component of the
  optically detected outflow. Only a small fraction of the
  total mass loss is required for this X-ray emitting
  component. Our favoured model requires shock velocities between 400 and
  500~km~s$^{-1}$. For a density $>10^5$~cm$^{-3}$ all dimensions of the
  shock cooling zone are only a few AU, so even in
  optical observations this cannot be resolved.}  {This X-ray emission mechanism in outflows may also operate
  in other, less absorbed T Tauri stars, in addition to
  corona and accretion spots.}

\keywords{stars: formation -- stars: winds/outflows -- stars:
  individual: DG Tau -- stars: mass-loss} \maketitle

\defcitealias{2008A&A...478..797G}{GS08}

\section{Introduction}
Jets and outflows seem to be a natural consequence of disc accretion, as they are common phenomena in star formation. Jets are an especially active area of study at present. Their innermost region is particularly interesting because it arises from the deepest in the potential well and thus probes the conditions there, which cannot be resolved by direct imaging. This innermost region is crucial for our understanding of jet launching and collimation. We need to constrain the different theoretical models of stellar winds \citep{1988ApJ...332L..41K,2005ApJ...632L.135M}, X-winds \citep{1994ApJ...429..781S} and disc winds \citep{1982MNRAS.199..883B,2005ApJ...630..945A} to produce reliable estimates of the associated angular momentum loss from the disc and possibly the star. This spin-down is needed to explain the slow rotation of accreting classical T Tauri stars (CTTS).

CTTS also have long been known as copious emitters of X-rays \citep{1999ARA&A..37..363F}, and we now recognise
that multiple emission mechanisms contribute to the observed radiation. Studies of large statistical samples clearly show a coronal component, which occasionally breaks out in hard X-ray flares \citep{2005ApJS..160..401P,2007A&A...468..463S}. Additionally a hot spot on the stellar surface heated by ongoing accretion contributes \citep{acc_model}. It shows up in X-ray line ratios originating in high-density regions, much denser than any corona \citep{2002ApJ...567..434K,twhya,bptau,v4046,2007A&A...465L...5A}.
There is relatively recent evidence of X-rays from the outflows themselves, starting with \object{HH 2} \citep{2001Natur.413..708P} and \object{HH 154} \citep{2002A&A...386..204F,2003ApJ...584..843B,2006A&A...450L..17F}.
X-rays very likely trace the fastest shocks in the jets. The first spatially extended X-ray emission of a CTTS jet was found using \emph{Chandra} by \citet{2005ApJ...626L..53G,2008A&A...478..797G} in \object{DG Tau}. This star shows a
spatially resolved, soft emission extending out to 5\arcsec{}. It is
not compatible with a point source origin, and \citet[][hereafter GS08]{2008A&A...478..797G} model it by a
combination of a linear source plus a Gaussian of $\approx140$~AU
diameter at the maximum distance to the star. 
Also, \citet{2007A&A...468..515G} find unusual spectral shapes in some CTTS distinguishing these so-called TAX (two absorber X-ray) sources from most other CTTS. Their spectra can be fitted well with a highly
absorbed hard component and a much less absorbed soft component.
They suggest that the marginally resolved soft excess is emission from
the base of the jet \citep[see also][]{Schneider}. In this paper we develop this picture further
in order to use the X-rays as a probe of the innermost, fastest
outflow and to constrain the jet's spatial extent and mass outflow
rate. This idea is supported by the finding of \citet{Schneider} that 
there is a small but significant spatial offset between the soft and the hard central components in DG~Tau,
so even the central component is marginally resolved. 

We focus specifically on DG~Tau, since the source is studied at many wavelengths, especially 
the stronger jet to the SW, but much less information is available for the fainter counter-jet to the NE.
In the optical the jet has been imaged with \emph{HST/STIS} several times, revealing not only jet rotation
\citep{2007ApJ...663..350C}, but also the presence of components with
different velocities in the approaching jet with up to 600 km~s$^{-1}$
\citep{2000ApJ...537L..49B}, after correcting for projection effects.

Based on analysis of forbidden lines, it was inferred that the jet
density rapidly falls off with increasing distance from the star
\citep{2000A&A...356L..41L,2000ApJ...537L..49B}. The jet extends to
0.5~pc as \object{HH 702} \citep{2007A&A...467.1197M}, where
shock-excited knots are clearly seen, but also closer to the star
there has to be a heating source to sustain the jet emission. Heating
by ambient shocks within the material itself is a good
candidate. \citet{2004A&A...416..213T} found a cool and slow molecular
wind with an opening angle near $90\deg$, which surrounds the faster
and hotter atomic outflow \citep[see also][]{2008ApJ...676..472B}.


\citet{2001A&A...367..959R} presented a 3D hydrodynamical simulation of the DG~Tau jet. They assume a sinusoidal velocity variation and precession. The precession broadens up the working surfaces in an otherwise well-collimated beam and produces several shocks where the faster material catches up with the previously ejected slower matter. In this model the density decreases naturally with the distance to the stellar source.

In this paper we examine the emission measure, column density and
temperature of the X-ray emitting region through spectral fitting
(Sect.~\ref{observations}) and interpret the soft component as a
tracer of the fastest component of the outflow (Sect.~\ref{model}). We
then discuss constraints from other wavelength bands and their
implication for the outflow origin in Sect.~\ref{discussion}. We
present our conclusions in Sect.~\ref{conclusion} and give a short
summary in Sect.~\ref{summary}.

\section{X-ray observations}
\label{observations}

\begin{figure}
\resizebox{\hsize}{!}{\includegraphics[angle=-90]{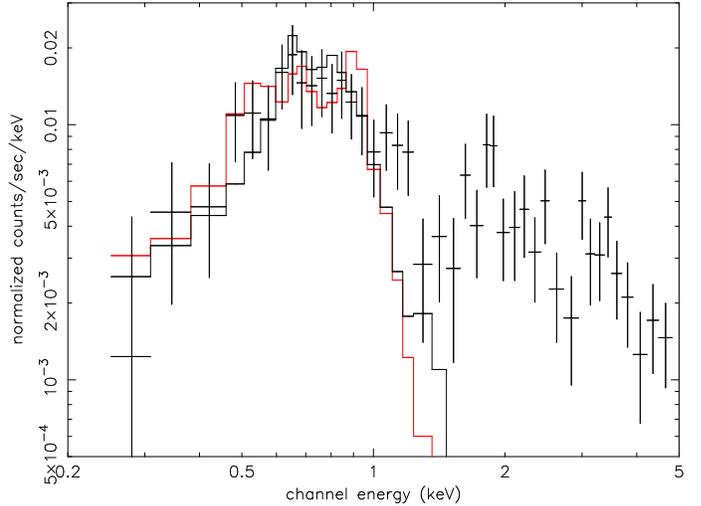}}
\caption{\label{spectrum}\emph{XMM-Newton} EPIC/PN data of
  DG~Tau. Overplotted are the contributions by the soft component for
  two different scenarios: $\textnormal{k}T=0.3$~keV ( $N_{\mathrm{H}}=2.3\times10^{21}$~cm$^{-2}$; black line) and
  $\textnormal{k}T=0.08$~keV ($N_{\mathrm{H}}=11\times10^{21}$~cm$^{-2}$; red/grey line).}
\end{figure}

We rely on the X-ray data presented by \citetalias{2008A&A...478..797G}, consisting of a 40~ks
\emph{XMM-Newton} observation (ObsID 0203540201) and four \emph{Chandra}
exposures (ObsIDs 4487, 6409, 7247 and 7246), which sum up to 90~ks
observation time. We retrieved this data from the archives, reduced it
using standard SAS and CIAO tasks, and analyse it in XSPEC. Because of
the low count rate we merge all available \emph{Chandra}
data. \citetalias{2008A&A...478..797G} show that the error bars on the
fitted values overlap between the four exposures for the soft
component. The luminosity in the hard component varies by a factor of
three, but the other parameters are compatible. Since we
  are most interested in modelling the soft component, variations in
  the luminosity and temperature of the hard component are not
  important here. We simultaneously fit the merged \emph{Chandra} and
the \emph{XMM-Newton} data with two thermal components with individual
absorption. We keep the parameters for \emph{Chandra} and
\emph{XMM-Newton} coupled in the soft component and leave the
temperature and normalisation between the hard components free,
because it is observed to be much stronger in the \emph{XMM-Newton}
observation due to intrinsic variability. As an illustration we show the data recorded by the EPIC/PN camera which covers the broadest energy range in
Fig.~\ref{spectrum}. As black and red/grey lines we show two fits to the soft
components with different temperatures. The fits shown use solar abundances
according to \citet{1998SSRv...85..161G}.

Due to the low count numbers, the statistical uncertainties are large.
In Fig.~\ref{spectrum} the black line shows a model with
$\textnormal{k}T=0.3$~keV and moderate absorption, very similar to
that presented in \citetalias{2008A&A...478..797G}. However, a model
with a much cooler temperature ($\textnormal{k}T=0.08$~keV), higher emission measure and larger
absorbing column density reproduces the data nearly as well, as
illustrated by the red/grey line in the figure. Thus, for the
  soft component, there is an ambiguity that makes it difficult to
  precisely determine the temperature, emission measure, and
  absorption column.

\begin{figure}
\resizebox{\hsize}{!}{\includegraphics{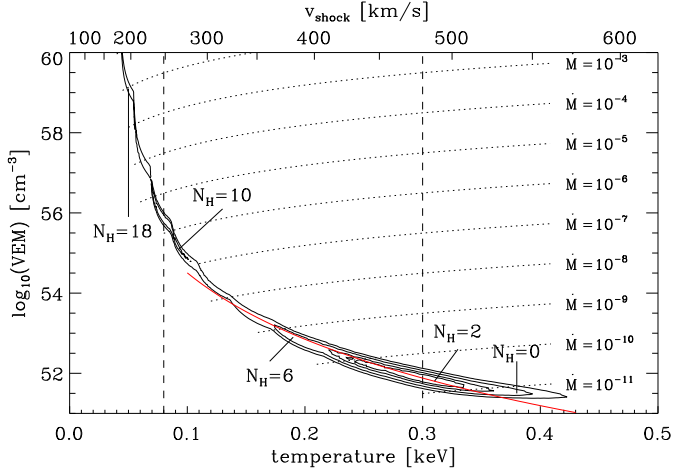}}
\caption{\label{vemkt}This shows confidence contours (at 68\,\%,
  90\,\%, 99\,\% and 99.95\,\% confidence level) fitting the VEM and
  the temperature of the soft component. The parameters in the
  hard component are held fixed. Assuming a simple shock model (see
  text for details) the shock velocity is given on the upper
  horizontal axis and mass loss rates in M$_{\sun}$~yr$^{-1}$ are
  marked in dotted lines in the plot. For some models absorbing column
  densities $N_{\mathrm{H}}$ are given in units of $10^{21}$~cm$^{-2}$. 
  The temperatures of the two models shown in Fig.~\ref{spectrum} are marked with
  dashed lines. Our fit from Eq.~\ref{eqnvemkt} is indicated with a red/grey line.}
\end{figure}

To quantify this, we have run a grid of models, keeping all parameters
for the hard component fixed and leaving the cool component's volume
emission measure (VEM hereafter), temperature and absorption column density free. The
resulting $\chi^2$-distribution is shown in Fig.~\ref{vemkt}. At the
99.95\,\% confidence level the distribution ranges over ten orders of
magnitude in emission measure and one order of magnitude in
temperature,  following a narrow valley in Temperature--VEM
  space. The most likely temperature region is 0.2-0.4~keV with a
second region within the 99\,\% confidence contour around 0.08~keV. 
 The secondary peak around 0.08 keV is caused mostly by \emph{Chandra} data; it
does not appear in a fit to the data from \emph{XMM-Newton}, which is more
sensitive to low energy X-rays than \emph{Chandra}.

  We verified that the position of this valley in $\chi^2$-space does
not change if the parameters of the hard component are left free.
  We fit the thermal
emission with two APEC models.  When using MEKAL, the minimum
  $\chi^2$-valley is in the same place, but the absolute minimum
  $\chi^2$ model is shifted to lower energies by up to 0.05~keV along
the valley. We regard this as an estimate for the systematic
uncertainty in the fit process.

 Due to this and the ambiguity in fitting
  the temperature and VEM of the soft component, rather than taking
  the formal minimum $\chi^2$ model, we consider a range of parameter
  space that is statistically allowed by the observations.  Future
  observations will be needed to pin down more precise values.  Thus,
  we fit a simple power-law to the minimum $\chi^2$-valley of figure
  \ref{vemkt}. This gives
\begin{equation}\label{eqnvemkt}
\frac{\mathrm{VEM}}{10^{52}\mathrm{ cm}^{-3}} \approx 0.45 \cdot \left(\frac{0.33\textnormal{ keV}}{\mathrm{k}T}\right)^{5.5}.
\end{equation}
The emission measure of the minimum $\chi^2$-valley deviates
less than a factor of 2 from equation
  (\ref{eqnvemkt}) in the energy range 0.1-0.4~keV (Fig.~\ref{vemkt}). We will use this relationship in the
  following section.

To get a handle on the physical conditions in the emitting region
further constraints are needed.  The fact that the central component of the soft emission
  does not deviate significantly from a point source \citep{Schneider}, gives an upper limit
on the emitting volume and therefore a lower limit on the density. Its size 
 should be less than 
  1\arcsec, corresponding to 140~AU at the distance of DG~Tau. At the
same time \citet{Schneider} found the soft component to be significantly
offset from the hard component at the stellar position by about
30~AU to the SW, the direction of the X-ray emitting, approaching jet.

\section{A wind model}
\label{model}

\begin{figure}
\resizebox{\hsize}{!}{\includegraphics{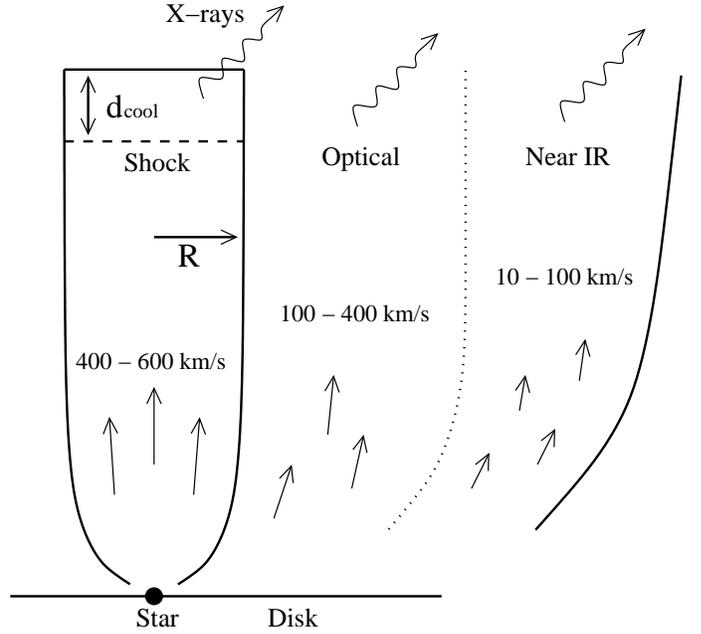}}
\caption{\label{cartoon}Sketch of the geometry. The flow launched
    from the region closest to the star travels with the highest
    velocity, and internal shocks in the flow there can reach the
    highest temperatures.}
\end{figure}

We use a simple, analytical model for the outflow. We discuss the observations giving rise to this model in detail in Sect.~\ref{otherobs}. Those observations indicate a collimated inner jet, which moves relatively fast. It is surrounded by slower and consecutively cooler components \citep[e.g.][see Sect.~\ref{otherobs} for more references]{2000ApJ...537L..49B}. This situation -- resembling the layers of an onion -- is sketched in Fig.~\ref{cartoon}. 
The outer wind is launched from the disc whereas the origin of the inner jets could be either the inner disc region or the star itself. Disc winds are launched at a temperatures of $\sim10^4$ K or much
less \citep{2002ApJ...566.1100A,2006ApJS..165..256H}.  Stellar winds, even if launched at coronal
temperatures, will cool adiabatically, and the radiative cooling time
can also be very short \citep{2007IAUS..243..299M}.  Therefore, 
beyond several tens of stellar radii, we expect all components of the outflow to be
too cool to emit X-rays. The outer layer contains molecular hydrogen and 
radiates predominantly in the IR; the more collimated and faster flows can be observed in the optical. 

\citet{2001A&A...367..959R} modelled successfully the 
optical observations with a jet model using a time-variable outflow speed. This gives rise to strong shocks, when faster gas catches up with previous ejecta. We assume that in the innermost outflow region the velocity of the flow is high enough to heat the post-shock material to X-ray emitting temperatures. The shock forms sufficiently far from the star that the flow is collimated.

Using the strong shock conditions we can transform the thermal energy
k$T$, where k is Boltzmann's constant and $T$ is the temperature, into
the pre-shock velocity in the shock rest frame $v_{\mathrm{shock}}$:
\begin{equation} 
\left(\frac{v_{\mathrm{shock}}}{500\textnormal{ km s}^{-1}}\right)^2 \approx 
\frac{\textnormal{k}T}{0.33\textnormal{ keV}} \; .
\label{vkt} 
\end{equation}
This velocity is given on the upper horizontal axis in
Fig.~\ref{vemkt}. We emphasise that, in the case of a travelling shock wave, the pre-shock gas velocity is the sum of $v_{\mathrm{shock}}$ and the motion of the shock front.
\citet{2002ApJ...576L.149R} provided a
  semi-analytical formulation for estimating the cooling length of gas
  heated in fast shocks ($\Delta v \ga 200$ km/s).  We checked this
analytic relation with more detailed simulations of the shock cooling
zone using our own code \citep{acc_model} and found that they agree
within a factor of a few. Thus, for the semi-analytic treatment
  here, we will use the formula of \citet{2002ApJ...576L.149R}, which can
  be written
\begin{equation}
d_{\mathrm{cool}} \approx 20.9 \mathrm{ AU} 
    \left(\frac{10^5\mathrm{ cm}^{-3}}{n_0}\right) 
    \left(\frac{v_{\mathrm{shock}}}{500\textnormal{ km s}^{-1}}\right)^{4.5},
\label{raga} 
\end{equation}
where $n_0$ is the pre-shock particle number density, equal to a
quarter of the post-shock number density $n$ in the strong shock
approximation.  

We assume an idealised cylindrical geometry for the post-shock
  flow, where the length of the cylinder is $d_{\mathrm{cool}}$. The
mass flow rate $\dot M_{\rm shock}$ through the shock front with
  area $A$ is then given by
\begin{eqnarray}
\label{eqn_mdot1}
\dot M_{\rm shock} & = & Av_{\mathrm{shock}}\rho_0,
\end{eqnarray}
where $\rho_0=\mu m_{\mathrm{H}}n_{\mathrm{ion}_0}$ is the pre-shock mass density
of the flow, $n_{\mathrm{ion}_0}$ is the initial number density of heavy particles, i.e. atoms and ions, $\mu$ the mean relative ion particle weight, and
$m_{\mathrm{H}}$ the mass of hydrogen. Assuming solar abundances, we adopt
$\mu=1.26$.  It is worth noting that $\dot M_{\rm shock}$ is less than the
  total mass loss rate in the flow, since $v_{\mathrm{shock}}$ is
  measured in the rest frame of the shock, and shocks in these systems
  are observed to travel at speeds of tens to hundreds of km~s$^{-1}$.

The area of the shock can be determined from
\begin{eqnarray}
\label{eqn_area}
A = \pi R^2 = {\mathrm{Volume} \over d_{\mathrm{cool}}} = 
    {VEM \over n_{\mathrm{e}} n_{\mathrm{ion}} d_{\mathrm{cool}}},
\end{eqnarray}
where $n_{\mathrm{e}}$ and $n_{\mathrm{ion}}$ are the post-shock
  electron and ion number density. Combing Eqs.
  \ref{vkt}--\ref{eqn_area}, and assuming $n_{\mathrm{ion}} = 4
  n_{\mathrm{ion}_0} = 0.83 n_{\mathrm{e}}$, the mass flow rate can be written
\begin{eqnarray}
\dot M_{\rm shock} & \approx & 2.7\cdot 10^{-11}\frac{M_{\sun}}{\textnormal{yr}} 
              \left(\frac{VEM}{10^{52}\textnormal{ cm}^{-3}}\right)
              \left(\frac{0.33\textnormal{ keV}}{\textnormal{k}T}\right)^{1.75} \; .
\label{mdot1}
\end{eqnarray}
It is convenient that, by adopting a simple formula for the
  cooling length (Eq.~\ref{raga}) and a simple shock geometry, the
  determination of the mass loss rate is independent of the density of
  the flow.

 We can now apply this specifically to the soft X-ray component of
  DG Tau.  By combining the observed relationship of Eq.~\ref{eqnvemkt} with Eq.~\ref{mdot1}, one obtains
\begin{eqnarray}
\dot M_{\rm shock} & \approx & 1.2\times10^{-11} \frac{M_{\sun}}{\textnormal{yr}}
          \left(\frac{0.33\textnormal{ keV}}{\textnormal{k}T}\right)^{7.25},
\label{mdot2}
\end{eqnarray}
or, using equation (\ref{vkt}),
\begin{eqnarray}
\dot M_{\rm shock} & \approx & 1.2\times10^{-11} \frac{M_{\sun}}{\textnormal{yr}}
          \left(\frac{500\textnormal{ km s}^{-1}}{v_{\mathrm{shock}}}\right)^{14.5}.
\label{mdot3}
\end{eqnarray}
These formulae indicate that the mass loss rate required to
explain the observed soft X-rays has a very steep dependence on the
observed temperature. It is clear that a better determination
of the mass loss rate requires a more precise temperature measurement.
In Fig.~\ref{vemkt} lines of constant
mass loss rate are marked as dotted lines according to
Eq.~\ref{mdot1}. Over the full length of the confidence contour the
 mass loss varies by ten orders of magnitude. 

\begin{figure}
\resizebox{\hsize}{!}{\includegraphics{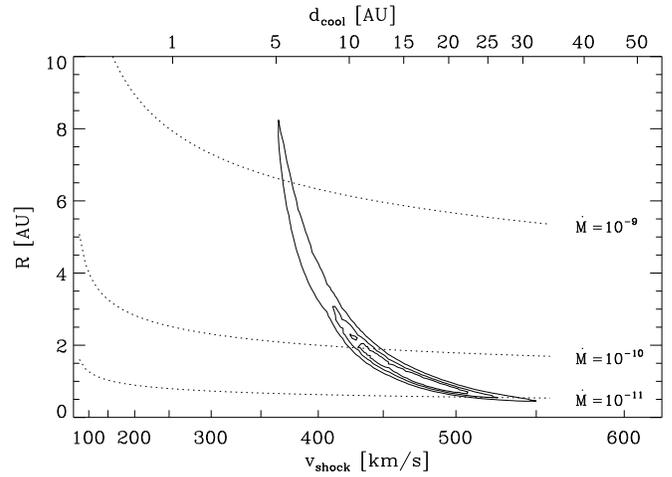}}
\caption{\label{rkt}Confidence contours (at 68\,\%, 90\,\% and 99\,\%
  confidence level) are shown for the parameters radius $R$ and shock
  velocity $v_{\mathrm{shock}}$ for the case of a cylindrical cooling
  zone at a fixed density of $n=10^5$~cm$^{-3}$. Dotted lines mark the mass loss
  rates in M$_{\sun}$~yr$^{-1}$. }
\end{figure}

We now look at the physical size of the shock region, $R$, which is
the radius of the base of the emitting cylinder of shocked
  gas. From Eqs. (\ref{vkt}), (\ref{raga}) and
  (\ref{eqn_area}), we can solve for the radius (in AU),
\begin{eqnarray}
R & \approx & 0.97 
        \left(\frac{10^5 \mathrm{cm}^{-3}}{n}\right)^{0.5}
        \left(\frac{VEM}{10^{52} \textnormal{ cm}^{-3}}\right)^{0.5} 
        \left(\frac{0.33 \mathrm{ keV}}{\mathrm{k}T}\right)^{1.125},
\label{eqnr}
\end{eqnarray}
and using the observed relationship for DG Tau (Eq.\ \ref{eqnvemkt}),
\begin{eqnarray}
R & \approx & 0.65 \textnormal{ AU}
              \left(\frac{10^5 \mathrm{cm}^{-3}}{n}\right)^{0.5} 
              \left(\frac{0.33 \mathrm{ keV}}{\mathrm{k}T}\right)^{3.875}.
\label{eqn_r2}
\end{eqnarray}
In Fig.~\ref{rkt} the confidence contours for the lower mass loss
rates are shown as functions of the radius of the shock at the
cylinder base and the cooling length assuming a density of
$n_{\mathrm{ion}}=10^5$~cm$^{-3}$ as motivated from the optical
observations. For other densites the scalings are $R \sim
\sqrt{\frac{1}{n}}$ (Eq.~\ref{eqnr}) and $d_{\mathrm{cool}}\sim
\frac{1}{n}$ (Eq.~\ref{raga}).

\section{Discussion}
\label{discussion}
In this section we will use the formulas and condition derived above to place limits on the physical mechanism producing the observed X-ray radiation.

\subsection{Elemental abundance}
The signal-to-noise ratio is not sufficient to fit elemental abundances, so we performed additional fits with the abundances found by \citet{2007A&A...473..589S} for the Taurus molecular cloud. As usual there is a strong degeneracy between emission measure and metallicity. On average reduced abundances can be compensated for by a larger emission measure. Changes in the composition, that is particularly the Ne/O and Ne/Fe ratio, lead to different thermal structures. The confidence contours for fits with the abundances from \citet{2007A&A...473..589S} follow the same valley in $\chi^2$-space, but are displaced about 0.07~keV to higher energies. Models below 0.27~keV are excluded on the 99\% confidence level.

It is not clear, however, if the coronal abundance is applicable to the jet, since the launching region of the X-ray
emitting, fast jet component is unknown.

\subsection{Limits on the size of the shock}
\label{sizeshock}
Figure~\ref{vemkt} shows two distinct regions of the parameter space within the 99\%~confidence level. That region in the upper left corner corresponds to a scenario of a large emission measure, which is heavily absorbed. The temperature is only around 0.08~keV and therefore the $v_{\mathrm{shock}}\approx250$~km~s$^{-1}$. As an upper limit we assume a density of $10^5$~cm$^{-3}$ (see next section). The cooling length according to Eq.~\ref{raga} should be about 1~AU. In a simple cylindrical geometry the volume emission measure, transformed to a
cylindrical radius with Eq.~\ref{eqnr}, requires a base shock with a radius of 1000~AU, but this would appear as an extended source, which is not observed for the central component (\citetalias{2008A&A...478..797G}, \citealt{Schneider}). Only a density above $10^7$~cm$^{-3}$ could push all shock dimensions below $\approx100$~AU, so they appear as point source in the X-ray observation. 

We deem this unreasonably high, also the low energy region is excluded in fits to the \emph{XMM-Newton} data alone or in models with different -- but reasonable -- abundances, so we concentrate on the less absorbed case with a higher $v_{\mathrm{shock}}$ in Fig.~\ref{rkt}. This parameter region can be realised with emission from the inner component of the jet only.


\subsection{Limits placed by other observations}
\label{otherobs}
Observations with the \emph{HST} allow to spatially resolve the
outflow of DG~Tau far below 1\arcsec{} from the central source. The
forbidden optical lines observed place density
constraints. \citet{2000ApJ...537L..49B} conducted seven exposures
with a long slit parallel to DG~Tau's outflow in H$\alpha$ and
forbidden optical lines. They analyse the line shift within 0\farcs5
from the star and find that the faster moving gas is more confined
towards the central axis. The emission of gas faster than
200~km~s$^{-1}$ is mostly confined to the inner slit, corresponding to
a cylinder of radius 15~AU or less. Comparison of intensity maps in
[\ion{O}{i}], [\ion{N}{ii}] and [\ion{S}{ii}] leads them to conclude
that the number density in the fast gas is higher than
$10^4$~cm$^{-3}$ (the line ratios are not sensitive to densities above this value). 
The lower velocity emission occupies a larger area
and originates in regions of lower densities of the order $10^3-10^4$~cm$^{-3}$. So, the inner, dense and fast components
seem to be surrounded by slower and less dense outflows, maybe
continuously down to a cool wind observed in H$_2$ with an opening
angle of 90\degr \citep{2004A&A...416..213T}.
Consistent density estimates of $10^5-10^6$~cm$^{-3}$ at distances
less than 0\farcs5, deprojected to $\approx100$~AU, were obtained by
\citet{2000A&A...356L..41L}. Strictly speaking, their line ratios give only lower limits on the density for the fastest innermost components, too, but in any case the outer layers do not contribute significantly to the X-ray radiation.

From the total line intensity \citet{1997A&A...327..671L} estimate a total mass loss rate
of $6.5\cdot 10^{-6}$~M$_{\sun}$~yr$^{-1}$ if the gas is heated by
shocks. With somewhat similar assumption \citet{1995ApJ...452..736H}
obtain $3\cdot 10^{-7}$~M$_{\sun}$~yr$^{-1}$. At a distance of 1\farcs2 from the central source \citet{2000A&A...356L..41L} estimate a mass loss rate of $1.4\cdot 10^{-8}$~M$_{\sun}$~yr$^{-1}$.
This evidence again favours the
lower corner in Fig.~\ref{vemkt}, shown in more detail in
Fig.~\ref{rkt}. 

In optical and IR line emission is detected blue-shifted up
to deprojected velocities of 600~km~s$^{-1}$
\citep{2000A&A...356L..41L,2000ApJ...537L..49B,2003ApJ...590..340P}.
An analysis of data covering more than a decade by \citet{2003ApJ...590..340P} indicates that the knots in the outflow move with a common proper motion of 0\farcs28~yr$^{-1}$, which translates into a deprojected velocity of 300~km~s$^{-1}$.

\subsection{The shocked gas component}
As shown in Sect.~\ref{sizeshock} the X-ray emission is unlikely to originate in the slowly moving components of the outflow. This leaves the innermost component as the probable emission origin. Unrealistically high densities are needed to produce an emission measure of $10^{56}$~cm$^{-3}$ for low shock speeds, so we favour a scenario where a strong shock with $v_{\mathrm{shock}}>400$~km~s$^{-1}$ heats up a relatively small component of the jet. In this case the emission measure is much smaller, so the required mass flux to power this emission is several orders of magnitude smaller than the total mass flux observed in the optical. The dotted lines in Fig.~\ref{rkt} show the mass loss depending on $v_{\mathrm{shock}}$. Taking into account the observed proper motion of the knots, the gas velocities of e.g. \citet{2000ApJ...537L..49B} are too low to explain the X-ray spectrum. Because less than $10^{-4}$ of the total flow has to reach the highest velocity to produce the X-rays, this small component was possibly not detected in the optical observations. In scenarios with standing (collimation shock) instead of travelling shock waves the innermost component is fast enough to explain the observed emission easily. 

After passing through the X-ray shock the material quickly radiates away its energy and cools down. Figure~\ref{rkt} shows all dimensions, the cooling length and the radius of the cylinder, to be small, therefore, they are not expected to be resolved in the \emph{HST} observations. The X-ray emitting shock is only a small disturbance and within a few AU the material has cooled down again to match the temperatures of the surrounding gas. The shocks we discuss here cannot be resolved in images directly, so only a spectral analysis can help to further refine the model.

\subsection{Formation of the shock front}
Although the \emph{Chandra} observations are
separated by nearly two years, no evidence of motion has been detected
\citep{Schneider}, but given the size of the error circles a velocity
$v_{\mathrm{front}}$ of the shock front of 100~km~s$^{-1}$ in the stellar rest frame cannot be
excluded. In this case the wind mass loss rate $\dot
M_{\mathrm{wind}}$ has to be calculated using the sum of
$v_{\mathrm{front}}$ and $v_{\mathrm{shock}}$:
\begin{equation}
\dot M_{\mathrm{wind}}=\frac{v_{\mathrm{front}}+v_{\mathrm{shock}}} {v_{\mathrm{shock}}} \dot M_{\mathrm{shock}}
\end{equation}
If these knots represent internal working surfaces the associated
$v_{\mathrm{shock}}$ is the difference between the gas motion and the
proper motion about 300~km~s$^{-1}$ for the knots.
This by itself is not sufficient to power X-ray emission in
the high temperature regime of Fig.~\ref{vemkt}.  The knots seem to be
launched irregularly, and the later \emph{Chandra} observations could
possibly probe a different knot than the earlier observations. A
relatively slow shock speed of 400~km~s$^{-1}$ corresponding to a
total mass loss rate of roughly $3 \times
10^{-10}$~M$_{\sun}$~yr$^{-1}$ (Fig.~\ref{rkt}) is then just
compatible with the highest optically detected outflow velocities. One
more scenario can be postulated: Instead of a single (stationary or
moving) shock front a higher number of small internal shocks could
produce the same X-ray signature. Each of these ``shocklets'' would
not show up in the optical observations, because radius and cooling
lengths were far below the current resolution limit. This requires a
clumpy outflow, where the jet contains a higher number of little knots
surrounded by the optically detected fast wind. Our calculated shock
area and cooling volume then represents the sum of all individual
shocks.
This argument is also valid, if the matter passes through multiple shock fronts and is reheated several times. The calculated mass flux rate still represents the summed flow of matter through all shock fronts, but it needs to be divided by the number of shocks to obtain the mass loss rate from the star. 
The optical observations discussed in Sect.~\ref{otherobs} do not resolve the postulated X-ray shock front, but the optically resolved shock fronts seem to emerge separated by 150~AU (A1 and A2 in \citet{2000ApJ...537L..49B}), so it appears reasonable that the X-rays may originate in a single shock front within 50 AU of the central star. 

From our modelling there is no clear distinction between a stationary
emission region as provided by a collimation shock or a moving
knot. Only further observations with a longer time baseline can
clarify this point. This is possible well within \emph{Chandra}'s lifetime. 

\subsection{The outflow origin}
The X-rays originate in the innermost and fastest
component, which may originate from the innermost part of the disc or
from the star itself. \citet{2007IAUS..243..299M} showed that a T
Tauri stellar wind with $\dot M_{\rm wind} \ga 10^{-11}$ M$_\odot$~yr$^{-1}$
cannot be coronally driven, but Alfv\'en wave driving may be possible \citep{2008arXiv0808.2250C}.
The spin rate of DG~Tau makes centrifugal launching of a stellar wind
unlikely, whereas the outer layers with lower flow velocities are most
likely centrifugally launched from the inner disc regions
\citep{2002ApJ...576..222B,2003ApJ...590L.107A}.

\subsection{Relation to extended X-ray jet of DG Tau}
The material heats up while passing through the shock front and we
could show that the cooling length is very small compared to spatial
extend of the jet. While cooling, the jet looses energy, so the strength of
the shock fronts decreases, their velocity jump diminishes with time.
In their paper \citet{2006A&A...449.1061D} show this in 
magneto-hydrodynamical models with a time variable ejection velocity leading to
travelling shock fronts.
Additionally, according to the optical observations, the density decreases with the
distance from the star. Therefore later heating events are less
luminous, so the resolved source at 5\arcsec{} is weaker. Although the resolved shock is a spatially extended source, this does not necessarily require multiple heating events between 2\arcsec{} and 5\arcsec{}, because the cooling length increases dramatically for low densities \citepalias{2008A&A...478..797G}.

If the cause of the shock is a collision with slower,
previously ejected matter we can expect this to happen more than
once. In this picture the knot of X-ray emission at 5\arcsec{} is
caused by matter catching up with an older ejection
event. \citetalias{2008A&A...478..797G} convincingly show that knots
even further out are far too faint in X-rays to be detected. 

\section{Conclusion}
\label{conclusion}

The soft X-ray emission of DG~Tau is consistent with our model of a
shock in the outflow. The cause of the shock is not yet clear, but
future observations can reveal if the position of the shock front is
stationary. In this case it is likely caused by the collimation of the
wind into the jet. If, on the other hand, the shock front moves, then
we can take this as indication that the outflow ejection
    velocity is highly variable. In both cases only the fastest
components of the optically detected outflow provide sufficient energy
to power the X-ray emission. Possible future grating observations
could greatly improve our estimate of the temperature in the emitting
region and narrow down the range of possible shock velocities and
therefore -- via Fig.~\ref{rkt} -- the mass loss rate.

DG~Tau is an exceptional CTTS in the sense that its central component
is so much absorbed that the origin of the soft emission cannot
coincide with the star. This makes it an ideal case to search for
alternative origins and our simple model successfully reproduces the
observation. On other CTTS, most notably \object{TW Hya}, detailed
models of an accretion shock explain the X-rays production very well
\citep{acc_model}. It seems possible, that TW~Hya also has a weak
contribution from wind shocks to its emission, but its outflows are
much weaker than those of DG~Tau, so the X-rays would be submerged in
the stellar emission. Next to accretion and coronal activity the wind
shocks, as discussed in this article, may hold as the third emission
mechanism of soft X-rays, although it is difficult to disentangle the
contributions except in special cases.


\section{Summary}
\label{summary}
In the X-ray spectrum of DG~Tau there is a spatially only marginally resolved soft
component. Because no grating information is available, the
temperature is only poorly constrained and plasma models with a wide
range of temperatures can reproduce the observations. The absorbing
column density is negligible in the case of plasma with a thermal
energy of $\textnormal{k}T=0.4$~keV and rises to $N_{\textnormal{H}} >
10^{22}$~cm$^{-2}$ for $\textnormal{k}T=0.1$~keV. For this second
scenario the fitted emission measure would be several orders of magnitude
higher. We reject this solution because it is not stable to small changes of the fit parameters and requires physically unreasonable values for the density in the shock region.

The emission can be explained by a shock and its corresponding cooling
zone in the innermost and fastest component of the optically detected
outflow. For densities $> 10^5$~cm$^{-3}$ this is consistent with all
available optical observations. Compared to the larger structures seen
in the forbidden optical lines, this X-ray cooling zone could be a
relatively thin layer. Only a small fraction of the total mass loss
is needed in the innermost layer in order to produce the observed luminosity.

\begin{acknowledgements}
HMG acknowledges support from DLR
under 50OR0105 and the Studienstiftung des deutschen Volkes. He thanks the University of Virginia
for hosting him while this work was done.
SPM is supported by the University of Virginia
through a Levinson/VITA Fellowship, partially funded by the Frank
Levinson Family Foundation through the Peninsula Community Foundation.  
ZYL is supported in part by AST-030768 and
NAG5-12102.
This research is based on observations obtained with XMM-Newton, an ESA science mission and Chandra, a NASA science mission.
\end{acknowledgements}

\bibliographystyle{aa} 
\bibliography{../articles}

\begin{thebibliography}{42}
\expandafter\ifx\csname natexlab\endcsname\relax\def\natexlab#1{#1}\fi

\bibitem[{{Anderson} {et~al.}(2003){Anderson}, {Li}, {Krasnopolsky}, \&
  {Blandford}}]{2003ApJ...590L.107A}
{Anderson}, J.~M., {Li}, Z.-Y., {Krasnopolsky}, R., \& {Blandford}, R.~D. 2003,
  \apjl, 590, L107

\bibitem[{{Anderson} {et~al.}(2005){Anderson}, {Li}, {Krasnopolsky}, \&
  {Blandford}}]{2005ApJ...630..945A}
{Anderson}, J.~M., {Li}, Z.-Y., {Krasnopolsky}, R., \& {Blandford}, R.~D. 2005,
  \apj, 630, 945

\bibitem[{{Ardila} {et~al.}(2002){Ardila}, {Basri}, {Walter}, {Valenti}, \&
  {Johns-Krull}}]{2002ApJ...566.1100A}
{Ardila}, D.~R., {Basri}, G., {Walter}, F.~M., {Valenti}, J.~A., \&
  {Johns-Krull}, C.~M. 2002, \apj, 566, 1100

\bibitem[{{Argiroffi} {et~al.}(2007){Argiroffi}, {Maggio}, \&
  {Peres}}]{2007A&A...465L...5A}
{Argiroffi}, C., {Maggio}, A., \& {Peres}, G. 2007, \aap, 465, L5

\bibitem[{{Bacciotti} {et~al.}(2000){Bacciotti}, {Mundt}, {Ray},
  {Eisl{\"o}ffel}, {Solf}, \& {Camezind}}]{2000ApJ...537L..49B}
{Bacciotti}, F., {Mundt}, R., {Ray}, T.~P., {et~al.} 2000, \apjl, 537, L49

\bibitem[{{Bacciotti} {et~al.}(2002){Bacciotti}, {Ray}, {Mundt},
  {Eisl{\"o}ffel}, \& {Solf}}]{2002ApJ...576..222B}
{Bacciotti}, F., {Ray}, T.~P., {Mundt}, R., {Eisl{\"o}ffel}, J., \& {Solf}, J.
  2002, \apj, 576, 222

\bibitem[{{Bally} {et~al.}(2003){Bally}, {Feigelson}, \&
  {Reipurth}}]{2003ApJ...584..843B}
{Bally}, J., {Feigelson}, E., \& {Reipurth}, B. 2003, \apj, 584, 843

\bibitem[{{Beck} {et~al.}(2008){Beck}, {McGregor}, {Takami}, \&
  {Pyo}}]{2008ApJ...676..472B}
{Beck}, T.~L., {McGregor}, P.~J., {Takami}, M., \& {Pyo}, T.-S. 2008, \apj,
  676, 472

\bibitem[{{Blandford} \& {Payne}(1982)}]{1982MNRAS.199..883B}
{Blandford}, R.~D. \& {Payne}, D.~G. 1982, \mnras, 199, 883

\bibitem[{{Coffey} {et~al.}(2007){Coffey}, {Bacciotti}, {Ray}, {Eisl{\"o}ffel},
  \& {Woitas}}]{2007ApJ...663..350C}
{Coffey}, D., {Bacciotti}, F., {Ray}, T.~P., {Eisl{\"o}ffel}, J., \& {Woitas},
  J. 2007, \apj, 663, 350

\bibitem[{{Cranmer}(2008)}]{2008arXiv0808.2250C}
{Cranmer}, S.~R. 2008, \apj, 689, {in press}

\bibitem[{{de Colle} \& {Raga}(2006)}]{2006A&A...449.1061D}
{de Colle}, F. \& {Raga}, A.~C. 2006, \aap, 449, 1061

\bibitem[{{Favata} {et~al.}(2006){Favata}, {Bonito}, {Micela}, {Fridlund},
  {Orlando}, {Sciortino}, \& {Peres}}]{2006A&A...450L..17F}
{Favata}, F., {Bonito}, R., {Micela}, G., {et~al.} 2006, \aap, 450, L17

\bibitem[{{Favata} {et~al.}(2002){Favata}, {Fridlund}, {Micela}, {Sciortino},
  \& {Kaas}}]{2002A&A...386..204F}
{Favata}, F., {Fridlund}, C.~V.~M., {Micela}, G., {Sciortino}, S., \& {Kaas},
  A.~A. 2002, \aap, 386, 204

\bibitem[{{Feigelson} \& {Montmerle}(1999)}]{1999ARA&A..37..363F}
{Feigelson}, E.~D. \& {Montmerle}, T. 1999, \araa, 37, 363

\bibitem[{{Grevesse} \& {Sauval}(1998)}]{1998SSRv...85..161G}
{Grevesse}, N. \& {Sauval}, A.~J. 1998, Space Science Reviews, 85, 161

\bibitem[{{G{\"u}del} {et~al.}(2008){G{\"u}del}, {Skinner}, {Audard}, {Briggs},
  \& {Cabrit}}]{2008A&A...478..797G}
{G{\"u}del}, M., {Skinner}, S.~L., {Audard}, M., {Briggs}, K.~R., \& {Cabrit},
  S. 2008, \aap, 478, 797

\bibitem[{{G{\"u}del} {et~al.}(2005){G{\"u}del}, {Skinner}, {Briggs}, {Audard},
  {Arzner}, \& {Telleschi}}]{2005ApJ...626L..53G}
{G{\"u}del}, M., {Skinner}, S.~L., {Briggs}, K.~R., {et~al.} 2005, \apjl, 626,
  L53

\bibitem[{{G{\"u}del} {et~al.}(2007){G{\"u}del}, {Telleschi}, {Audard},
  {L.~Skinner}, {Briggs}, {Palla}, \& {Dougados}}]{2007A&A...468..515G}
{G{\"u}del}, M., {Telleschi}, A., {Audard}, M., {et~al.} 2007, \aap, 468, 515

\bibitem[{{G{\"u}nther} {et~al.}(2006){G{\"u}nther}, {Liefke}, {Schmitt},
  {Robrade}, \& {Ness}}]{v4046}
{G{\"u}nther}, H.~M., {Liefke}, C., {Schmitt}, J.~H.~M.~M., {Robrade}, J., \&
  {Ness}, J.-U. 2006, \aap, 459, L29

\bibitem[{{G{\"u}nther} {et~al.}(2007){G{\"u}nther}, {Schmitt}, {Robrade}, \&
  {Liefke}}]{acc_model}
{G{\"u}nther}, H.~M., {Schmitt}, J.~H.~M.~M., {Robrade}, J., \& {Liefke}, C.
  2007, \aap, 466, 1111

\bibitem[{{Hartigan} {et~al.}(1995){Hartigan}, {Edwards}, \&
  {Ghandour}}]{1995ApJ...452..736H}
{Hartigan}, P., {Edwards}, S., \& {Ghandour}, L. 1995, \apj, 452, 736

\bibitem[{{Herczeg} {et~al.}(2006){Herczeg}, {Linsky}, {Walter}, {Gahm}, \&
  {Johns-Krull}}]{2006ApJS..165..256H}
{Herczeg}, G.~J., {Linsky}, J.~L., {Walter}, F.~M., {Gahm}, G.~F., \&
  {Johns-Krull}, C.~M. 2006, \apjs, 165, 256

\bibitem[{{Kastner} {et~al.}(2002){Kastner}, {Huenemoerder}, {Schulz},
  {Canizares}, \& {Weintraub}}]{2002ApJ...567..434K}
{Kastner}, J.~H., {Huenemoerder}, D.~P., {Schulz}, N.~S., {Canizares}, C.~R.,
  \& {Weintraub}, D.~A. 2002, \apj, 567, 434

\bibitem[{{Kwan} \& {Tademaru}(1988)}]{1988ApJ...332L..41K}
{Kwan}, J. \& {Tademaru}, E. 1988, \apjl, 332, L41

\bibitem[{{Lavalley} {et~al.}(1997){Lavalley}, {Cabrit}, {Dougados}, {Ferruit},
  \& {Bacon}}]{1997A&A...327..671L}
{Lavalley}, C., {Cabrit}, S., {Dougados}, C., {Ferruit}, P., \& {Bacon}, R.
  1997, \aap, 327, 671

\bibitem[{{Lavalley-Fouquet} {et~al.}(2000){Lavalley-Fouquet}, {Cabrit}, \&
  {Dougados}}]{2000A&A...356L..41L}
{Lavalley-Fouquet}, C., {Cabrit}, S., \& {Dougados}, C. 2000, \aap, 356, L41

\bibitem[{{Matt} \& {Pudritz}(2005)}]{2005ApJ...632L.135M}
{Matt}, S. \& {Pudritz}, R.~E. 2005, \apjl, 632, L135

\bibitem[{{Matt} \& {Pudritz}(2007)}]{2007IAUS..243..299M}
{Matt}, S. \& {Pudritz}, R.~E. 2007, in IAU Symposium, Vol. 243, IAU Symposium,
  ed. J.~{Bouvier} \& I.~{Appenzeller}, 299--306

\bibitem[{{McGroarty} {et~al.}(2007){McGroarty}, {Ray}, \&
  {Froebrich}}]{2007A&A...467.1197M}
{McGroarty}, F., {Ray}, T.~P., \& {Froebrich}, D. 2007, \aap, 467, 1197

\bibitem[{{Pravdo} {et~al.}(2001){Pravdo}, {Feigelson}, {Garmire}, {Maeda},
  {Tsuboi}, \& {Bally}}]{2001Natur.413..708P}
{Pravdo}, S.~H., {Feigelson}, E.~D., {Garmire}, G., {et~al.} 2001, \nat, 413,
  708

\bibitem[{{Preibisch} {et~al.}(2005){Preibisch}, {Kim}, {Favata}, {Feigelson},
  {Flaccomio}, {Getman}, {Micela}, {Sciortino}, {Stassun}, {Stelzer}, \&
  {Zinnecker}}]{2005ApJS..160..401P}
{Preibisch}, T., {Kim}, Y.-C., {Favata}, F., {et~al.} 2005, \apjs, 160, 401

\bibitem[{{Pyo} {et~al.}(2003){Pyo}, {Kobayashi}, {Hayashi}, {Terada}, {Goto},
  {Takami}, {Takato}, {Gaessler}, {Usuda}, {Yamashita}, {Tokunaga}, {Hayano},
  {Kamata}, {Iye}, \& {Minowa}}]{2003ApJ...590..340P}
{Pyo}, T.-S., {Kobayashi}, N., {Hayashi}, M., {et~al.} 2003, \apj, 590, 340

\bibitem[{{Raga} {et~al.}(2001){Raga}, {Cabrit}, {Dougados}, \&
  {Lavalley}}]{2001A&A...367..959R}
{Raga}, A., {Cabrit}, S., {Dougados}, C., \& {Lavalley}, C. 2001, \aap, 367,
  959

\bibitem[{{Raga} {et~al.}(2002){Raga}, {Noriega-Crespo}, \&
  {Vel{\'a}zquez}}]{2002ApJ...576L.149R}
{Raga}, A.~C., {Noriega-Crespo}, A., \& {Vel{\'a}zquez}, P.~F. 2002, \apjl,
  576, L149

\bibitem[{{Scelsi} {et~al.}(2007){Scelsi}, {Maggio}, {Micela}, {Briggs}, \&
  {G{\"u}del}}]{2007A&A...473..589S}
{Scelsi}, L., {Maggio}, A., {Micela}, G., {Briggs}, K., \& {G{\"u}del}, M.
  2007, \aap, 473, 589

\bibitem[{{Schmitt} {et~al.}(2005){Schmitt}, {Robrade}, {Ness}, {Favata}, \&
  {Stelzer}}]{bptau}
{Schmitt}, J.~H.~M.~M., {Robrade}, J., {Ness}, J.-U., {Favata}, F., \&
  {Stelzer}, B. 2005, \aap, 432, L35

\bibitem[{{Schneider} \& {Schmitt}(2008)}]{Schneider}
{Schneider}, P.~C. \& {Schmitt}, J.~H.~M.~M. 2008, \aap, 488, L13

\bibitem[{{Shu} {et~al.}(1994){Shu}, {Najita}, {Ostriker}, {Wilkin}, {Ruden},
  \& {Lizano}}]{1994ApJ...429..781S}
{Shu}, F., {Najita}, J., {Ostriker}, E., {et~al.} 1994, \apj, 429, 781

\bibitem[{{Stelzer} {et~al.}(2007){Stelzer}, {Flaccomio}, {Briggs}, {Micela},
  {Scelsi}, {Audard}, {Pillitteri}, \& {G{\"u}del}}]{2007A&A...468..463S}
{Stelzer}, B., {Flaccomio}, E., {Briggs}, K., {et~al.} 2007, \aap, 468, 463

\bibitem[{{Stelzer} \& {Schmitt}(2004)}]{twhya}
{Stelzer}, B. \& {Schmitt}, J.~H.~M.~M. 2004, \aap, 418, 687

\bibitem[{{Takami} {et~al.}(2004){Takami}, {Chrysostomou}, {Ray}, {Davis},
  {Dent}, {Bailey}, {Tamura}, \& {Terada}}]{2004A&A...416..213T}
{Takami}, M., {Chrysostomou}, A., {Ray}, T.~P., {et~al.} 2004, \aap, 416, 213

\end{thebibliography}
\end{document}